\documentclass[a4paper,fleqn,usenatbib]{mnras}

\DeclareTextFontCommand{\textbf}{\normalfont}

\usepackage[T1]{fontenc}
\usepackage{ae,aecompl}

\usepackage{graphicx}	
\usepackage{amsmath}	
\usepackage{amssymb}	
\usepackage{lscape}
\usepackage{afterpage}   
\usepackage{color}

\title[Eclipse timing of white dwarf binaries]{Long-term eclipse timing of white dwarf binaries: an observational hint of a magnetic mechanism at work}

\author[M.C.P. Bours et al.]{M.C.P. Bours$^{1,2}$\thanks{E-mail: madelon.bours@uv.cl},
   T.R. Marsh$^2$,
   S.G. Parsons$^1$,
   V.S. Dhillon$^{3,4}$,
   R.P. Ashley$^2$,
   \newauthor 
   J.P. Bento$^5$,
   E. Breedt$^2$,
   T. Butterley$^6$,
   C. Caceres$^{1,7}$,
   C.M. Copperwheat$^8$,
   \newauthor 
   L.K. Hardy$^3$,
   J.J. Hermes$^9$,
   P. Irawati$^{10}$,
   P. Kerry$^3$,
   D. Kilkenny$^{11}$,
   S.P. Littlefair$^3$,
   \newauthor 
   M.J. McAllister$^3$,
   S. Rattanasoon$^{3,10}$,
   D.I. Sahman$^3$,
   M. Vu\v ckovi\'c$^1$,
   R.W. Wilson$^6$ \\
   $^1$Departmento de F\'isico y Astronom\'ia, Universidad de Valpara\'iso, Avenida Gran Breta\~{n}a 1111, Valpara\'iso, Chile. \\
   $^2$Department of Physics, University of Warwick, Coventry CV4 7AL, UK. \\
   $^3$Department of Physics and Astronomy, University of Sheffield, Sheffield S3 7RH, UK. \\
   $^4$Instituto de Astrof\'isica de Canarias, E-38205 La Laguna, Santa Cruz de Tenerife, Spain. \\
   $^5$Research School of Astronomy and Astrophysics, Australian National University, Canberra, ACT 2611, Australia. \\
   $^6$Centre for Advanced Instrumentation, Department of Physics, University of Durham, South Road, Durham DH1 3LE, UK. \\
   $^7$Millennium Nucleus `Protoplanetary discs in ALMA Early Science', Universidad de Valparaiso, Valparaiso 2360102, Chile. \\
   $^8$Astrophysics Research Institute, Liverpool John Moores University, Twelve Quays House, Birkenhead, CH41 1LD, UK. \\
   $^9$Hubble Fellow - Department of Physics, University of North Carolina, Chapel Hill, NC 27599, USA. \\
   $^{10}$National Astronomical Research Institute of Thailand, 191 Siriphanich Bldg., Huay Kaew Road, Chiang Mai 50200, Thailand. \\
   $^{11}$Department of Physics, University of the Western Cape, Bellville 7535, South Africa. \\
}

\date{Accepted XXX. Received YYY; in original form ZZZ}

\pubyear{2016}

\begin{document}
\label{firstpage}
\pagerange{\pageref{firstpage}--\pageref{lastpage}}
\maketitle

\begin{abstract} 
We present a long-term programme for timing the eclipses of white dwarfs in close binaries to measure apparent and/or real variations in their orbital periods. Our programme includes 67 close binaries, both detached and semi-detached and with M-dwarfs, K-dwarfs, brown dwarfs or white dwarfs secondaries. In total, we have observed \textbf{more than} 650 white dwarf eclipses. We use this sample to search for orbital period variations and aim to identify the underlying cause of these variations. We find that the probability of observing orbital period variations increases significantly with the observational baseline. In particular, all binaries with baselines exceeding 10~yrs, \textbf{with secondaries of spectral type K2 -- M5.5, show variations in the eclipse arrival times that in most cases amount to several minutes. In addition, among those with baselines shorter than 10~yrs, binaries with late spectral type ($>$M6),} brown dwarf or white dwarf secondaries appear to show no orbital period variations. This is in agreement with the so-called Applegate mechanism, which proposes that magnetic cycles in the secondary stars can drive variability in the binary orbits. We also present new eclipse times of NN\,Ser, which are still compatible with the previously published circumbinary planetary system model, although only with the addition of a quadratic term to the ephemeris. Finally, we conclude that we are limited by the relatively short observational baseline for many of the binaries in the eclipse timing programme, and therefore cannot yet draw robust conclusions about the cause of orbital period variations in evolved, white dwarf binaries.
\end{abstract}




\section{Introduction}
The evolution of close, evolved binaries is determined by the binary's angular momentum and the stellar masses, and by how these parameters change with time. The stellar masses can change through mass-transfer between the stars or by mass-loss through a stellar wind, although the latter is usually very small compared to the star's mass. Considering white dwarf + low-mass main sequence stars, at short orbital periods ($\lesssim$~2h) the main change in the binary's angular momentum occurs through the emission of gravitational wave radiation \citep{Paczynski67, Faulkner71, LandauLifshitz75}. At longer orbital periods ($\gtrsim$~3h) the loss of angular momentum is driven by a mechanism called magnetic braking instead \citep{Huang66, Mestel68, Verbunt81}. This occurs because the main-sequence star emits a stellar wind that flows away from the star and is forced by the magnetic field to corotate with the star out to the Alv\'en radius. There, the matter decouples from the magnetic field and takes angular momentum with it, thereby decreasing the spin angular momentum of the star. This phenomenon has been measured indirectly by its effect on the rotation rate of single stars \citep{Schatzman62, Kraft67}. In close binaries, tides force the low-mass main sequence star to rotate synchronously with the orbital motion, so that the star's spin period equals the binary's orbital period. Therefore the angular momentum that is carried away by the stellar wind is effectively removed from the binary's orbital angular momentum, causing the binary's orbit to decrease over time. Magnetic braking is especially important in cataclysmic variable stars and their progenitors. These binaries have separations that are small enough for magnetic braking to drive the binaries closer together and thereby drive evolution of the ongoing mass-transfer and the binary itself \citep{Rappaport83, Knigge11}. 

In addition to these secular processes, other processes may be at work that make it appear as if the binary is losing angular momentum, while this is in fact not the case. The two most popular theories include one now often referred to as Applegate's mechanism (\citealt{Applegate87, Applegate92}, see also \citealt{Lanza98}), and the presence of circumbinary planet-like or brown dwarf-like bodies of mass \citep[see for example][]{Beavers86, Beuermann13, Marsh14}. 

\citet{Applegate92} proposed that a main sequence star in a close binary may experience magnetic cycles during which angular momentum is redistributed between the core and the outer layers of the star by the turbulent motion in the convective region and the torques produced by the differential rotation. This also causes the star to deform and become oblate, therefore changing its gravitational quadrupole moment. In turn, this couples to the binary's orbit, which subsequently changes its orbital period on the same quasi-periodic time scales as the magnetic activity cycles that drive this mechanism. Note that no angular momentum is lost from the binary during this process, it is simply redistributed within the main-sequence star. Nonetheless, this requires energy and since the orbital period variations are driven by the main-sequence star in the binary the maximum amplitude of the variations is determined by the maximum amount of energy available in this star. Generally, in close white dwarf binaries, the main-sequence companions are M-dwarfs. The luminosities of these stars can be considered as their maximum energy budget available to drive the magnetic cycles, their physical distortions, and therefore the binary's orbital period variations. For certain white dwarf + M-dwarf binaries in which large variations of this kind have been observed, the energy budget is seemingly insufficient \citep{Brinkworth06,Bours14b}. However, note that a modified version of Applegate's mechanism predicts that orbital period variations can be induced while only requiring a fraction of the energy of the original mechanism of \citet{Applegate92}, see \citet{Lanza98} and \citet{Lanza06}. 

Lastly, an unseen companion in a wide, circumbinary orbit around a white dwarf + M-dwarf binary may make it appear as if the binary itself is losing angular momentum and changing its orbital period. We will refer to such companions as circumbinary planets or circumbinary brown dwarfs hereafter. The presence of this extra mass causes the binary to periodically change its distance to a given observer, thereby causing eclipses to be observed slightly advanced or delayed with respect to the expected eclipse time in a sinusoidal manner \citep{Irwin59}. In the last few years the first circumbinary planets have been discovered around double main-sequence star binaries through transits in the light curves \citep{Doyle11, Welsh12, Orosz12b, Orosz12a}, leaving no doubt about the existence of planets in so-called P-type orbits \citep{Dvorak86}. However, in close white dwarf binaries the primary star has evolved off the main-sequence, and the binary has likely gone through a common-envelope phase. This may have destroyed any planetary system present \citep{Veras12, Mustill13}, leaving it far from certain that planetary systems exist around white dwarf + main-sequence star binaries. On the other hand, circumbinary planets may be able to form again in a second phase of planet formation, triggered in the ejecta of the binary's common-envelope phase \citep{Schleicher14, Bear14}, and so it is not completely unlikely that some indeed exist. However, the first direct search for a circumbinary brown dwarf, whose presence was suggested by substantial eclipse timing variations, has resulted in a non-detection \citep{Hardy15}.

All four processes previously mentioned may be measured in observational data of a binary if it is possible to measure a regular, unrelated phenomenon in that binary. This could be the eclipse of a white dwarf, hot subdwarf star or neutron star \citep{Wood63, Parsons10b, Kilkenny11, Hermes12b}, the very regular pulses that a magnetic neutron star emits \citep{Wolszczan92, Wolszczan94}, or pulsations of stars themselves \citep{Silvotti07, Mullally08, Hermes13a2}. Such a precise clock allows the observer to measure changes in the orbital or spin periods in the system. Generally, observational data would also allow one to distinguish between certain mechanisms through the time scale on which the phenomena occur. Both magnetic braking and gravitational wave emission are secular processes, evolving slowly and steadily and typically taking 10$^8$ - 10$^9$ years. Orbital period changes caused by an Applegate-like mechanism or apparent variations caused by circumbinary objects on the other hand typically take place on 10 - 100 year time scales.

To complement the ongoing discussion about observed apparent orbital period variations in close binaries, we have set up an eclipse timing programme that focuses on measuring eclipse times of a large number of such binaries. In this paper we focus on close binaries in which the primary star is a white dwarf. This offers the advantages that the eclipse ingress and egress features are short and sharp. In addition, the white dwarfs are often substantially hotter than their low-mass companions, leading to deep eclipses. Both advantages facilitate precise and accurate timing of the eclipses. Through regular eclipse observations over a baseline of years to decades, we hope to create a picture that shows which of these evolved binaries display orbital period variations and how prominently these are present in certain categories of binaries. Large surveys such as the Sloan Digital Sky Survey \citep[SDSS,][]{York00} and the Catalina Sky Survey \citep[CSS,][]{Drake09} have increased the number of known eclipsing white dwarf binaries to several hundred in recent years \citep{Ritter03,Parsons13,Parsons15}. With this increase in sample size, it is now possible to perform long-term monitoring of an entire population of evolved binaries. Previous studies using smaller samples of binaries have already shown that almost all binaries that have been monitored for more than $\sim$~5~years show apparent orbital period variations \citep{Zorotovic13}.


\section{Observations}
\subsection{Time scales used for timing periodic phenomena}
Due to the Earth's motion around the Sun and the finite speed of light, the exact time that a certain event is observed depends upon the changing position of the observer, as well as the particular clock used to express the time. To be able to compare observations from different epochs, the observed time can be converted to a number of time stamps \citep{Eastman10}. To obtain the highest precision in this paper, we use Barycentric Modified Julian Dates (BMJD), which are corrected for the motion of the Sun around the barycentre of the Solar system. Together with Barycentric Dynamical Time, which is a relativistic time standard in the reference frame of the Solar system barycentre, all times in this paper are quoted in BMJD(TDB). For data taken as part of the eclipse timing programme described here, we also include the MJD(UTC) eclipse times, which is the time scale used to time stamp the observations. Times taken from the literature are, when not already in BMJD(TDB), converted to this time scale \citep{Eastman10}.

\subsection{Targets}
Our eclipse timing programme includes 67 eclipsing binaries. Of these, 58 are detached binaries, with 54 white dwarf + main-sequence star binaries, 1 white dwarf + brown dwarf binary and 3 double white dwarf binaries. Basic information about these targets is listed in Table~\ref{tab:detachedtargets}. 

The remaining 9 targets are semi-detached white dwarf binaries, also known as cataclysmic variables. There are 3 cataclysmic variables in our programme that have strongly magnetic white dwarfs, and are therefore classified as polars. In addition, there is 1 cataclysmic variable with a brown dwarf donor star. See Table~\ref{tab:cvtargets} for details of these semi-detached binaries. Although the erratic features in the light curves caused by the variable accretion rate in these systems complicate the determination of accurate eclipse times, their tendency to experience outbursts also means that some of these cataclysmic variables have first been discovered decades ago. Their eclipse observations therefore tend to span a much longer baseline than those of the more-recently discovered detached binaries. This may be useful for revealing periodic variability of the eclipse times on decade time scales.

\subsection{Observing strategy}
The eclipse observations presented here were taken using a number of telescopes and instruments. Mostly they were done with ULTRACAM \citep{Dhillon07} on the \textbf{4.2m} William Herschel Telescope (WHT) and the \textbf{3.6m} New Technology Telescope (NTT), ULTRASPEC \citep{Dhillon14} on the \textbf{2.4m} Thai National Telescope (TNT) and RISE on the \textbf{2.0m} Liverpool Telescope \citep[LT;][]{Steele04, Steele08}. A full list of the telescopes and instruments used to obtain our observations can be found in Table~\ref{tab:telescopes}. For each binary in the timing programme, the number of new eclipse times are listed in Tables~\ref{tab:detachedtargets} and~\ref{tab:cvtargets}. The eclipse times themselves will be available online (through Vizier\footnote{http://vizier.u-strasbg.fr/viz-bin/VizieR}) for each binary. There we also include eclipse times from the literature, where available, for the purpose of completeness. Note though, that we exclude eclipse times with uncertainties exceeding 20~s.

In total, the sample includes more than 650 new, previously unpublished eclipse times, ranging from 1 for some of the newer or very long-period white dwarf binaries up to 20 -- 30 for those binaries that were starting to show O-C variability and therefore justified close monitoring. 

{ \footnotesize
\makeatletter
\begin{table*}
\caption{List of the 58 detached eclipsing white dwarf binaries included in the monitoring programme described in this paper, sorted by RA. The SIMBAD identifier, frequently used alternative name and spectral type of the companion star to the white dwarf are given where available. The numbers in the parentheses following the zero-point and orbital period of the best linear ephemeris indicate the uncertainty in the last digits. The column marked with a \# shows the number of new eclipse times presented here and the last column lists references for the discovery, a detailed study of the binary and/or the companion's spectral type. }
\label{tab:detachedtargets} 
\begin{tabular}{llllllll}
\hline 
ID & \multicolumn{2}{c}{white dwarf binary}  &  SpT$_2$       &  \multicolumn{2}{c}{best linear ephemeris}                 & \#  &  ref \\
   & SIMBAD identifier & alternative         &                &  T$_0$ (BMJD$_{\mathrm{TDB}}$) & P$_{\mathrm{orb}}$ (days) &     &      \\
\hline 
 1 & SDSS J002412.87+174531.4  &  -                 &  M3         &  56482.1968(1)        &  0.20003847(2)        &   4  &  1,2 \\
 2 & SDSS J010623.01-001456.2  &  -                 &  M6         &  55059.056123(6)      &  0.0850153291(5)      &  28  &  3,2 \\
 3 & SDSS J011009.09+132616.3  &  -                 &  M4         &  53993.94904(3)       &  0.332686773(4)       &  11  &  4,5 \\
 4 & SDSS J013851.49-001621.6  &  -                 &  M5         &  55867.00740(1)       &  0.0727649720(9)      &   2  &  6,5 \\
 5 & PTF1 J015256.60+384413.4  &  PTFEB28.235       &  M3         &  56195.16844(2)       &  0.38612034(2)        &   6  &  7 \\
 6 & SDSS J025953.32-004400.2  &  -                 &  M3         &  51819.4150(10)  &  0.1441834(1)  &   1  &  8,9 \\
 7 & SDSS J030308.35+005444.1  &  -                 &  M4.5       &  53991.11730(1)       &  0.1344376678(8)      &  11  &  4,10,5 \\
 8 & SDSS J030856.55-005450.6  &  -                 &  M3         &  56210.15322(4)       &  0.185959516(9)       &   5  &  9,2 \\
 9 & WD 0312+019               &  -                 &  -          &  56195.206351(1)      &  0.305296762(2)       &  12  &  11,12 \\
10 & NLTT 11748                &  -                 &  WD         &  55772.041389(3)      &  0.235060481(1)       &  12  &  13,14 \\
11 & V471 Tau                  &  -                 &  K2         &  54027.9530(1)        &  0.521183431(8)       &   1  &  15 \\
12 & RR Cae                    &  -                 &  M4         &  51522.54847(4)       &  0.303703678(4)       &  22  &  16 \\
13 & SDSS J082145.27+455923.3  &  -                 &  M2         &  55989.03882(1)       &  0.509092021(6)       &   5  &  17,5 \\
14 & SDSS J083845.86+191416.5  &  CSS 40190         &  M5         &  53469.22016(3)       &  0.130112311(1)       &  14  &  18,5 \\
15 & SDSS J085746.18+034255.3  &  CSS 03170         &  M8         &  55552.7127639(9)     &  0.0650965384(1)      &  16  &  19 \\
16 & SDSS J090812.03+060421.2  &  CSS 080502        &  M4         &  53466.33450(4)       &  0.149438038(2)       &  26  &  18,5 \\
17 & SDSS J092741.73+332959.1  &  -                 &  M3         &  56074.90612(3)       &  2.30822561(9)        &   6  &  17,5 \\
18 & SDSS J093508.00+270049.2  &  -                 &  -          &  56602.8398(2)        &  0.2010331(1)         &   5  &  20 \\
19 & SDSS J093947.95+325807.3  &  CSS 38094         &  M4         &  55587.308823(6)      &  0.330989665(2)       &   6  &  18,5 \\
20 & SDSS J094634.49+203003.3  &  -                 &  M5         &  56032.94566(3)       &  0.252861432(9)       &   8  &  17,5 \\
21 & SDSS J095719.24+234240.7  &  CSS 41631         &  M2         &  55548.35703(5)       &  0.150870797(7)       &  13  &  18,5 \\
22 & SDSS J095737.59+300136.5  &  -                 &  M3         &  56014.975114(32)  &  1.9261248(12)  &   0  &  17,5 \\
23 & SDSS J100559.10+224932.2  &  CSS 41177         &  WD         &  55936.3446717(5)     &  0.1160154373(3)      &  13  &  18,21,22 \\
24 & SDSS J101356.32+272410.6  &  -                 &  M4         &  53831.12550(2)       &  0.1290403812(8)      &   7  &  1,5 \\
25 & SDSS J102102.25+174439.9  &  -                 &  M4         &  56664.88435(1)       &  0.140358755(5)       &   1  &  17,23,5 \\
26 & SDSS J102857.78+093129.8  &  -                 &  M2.5       &  56001.0950(4)        &  0.23502508(10)       &  13  &  17,5 \\
27 & SDSS J105756.93+130703.5  &  -                 &  M5         &  56010.0627(2)        &  0.12516213(4)        &   7  &  17,5 \\
28 & SDSS J112308.39-115559.2  &  -                 &  M3.5       &  56364.2935(5)  &  0.7691358(14)  &   1  &  1,2 \\
29 & SDSS J121010.13+334722.9  &  -                 &  M5         &  54923.03353(5)       &  0.124489790(4)       &  18  &  24,5 \\
30 & SDSS J121258.25-012310.1  &  -                 &  M4         &  54104.20945(5)       &  0.335870877(7)       &  11  &  25,26,5 \\
31 & SDSS J122339.61-005631.2  &  -                 &  M6         &  55707.016990(7)      &  0.0900780296(6)      &  10  &  27,17,5 \\
32 & SDSS J124432.25+101710.8  &  CSS 25601         &  M4         &  53466.36035(8)       &  0.227856372(5)       &   6  &  18,5 \\
33 & SDSS J130733.49+215636.7  &  -                 &  M4         &  56007.22121(6)       &  0.21632235(1)        &   9  &  17,5 \\
34 & SDSS J132518.18+233808.0  &  CSS 21616         &  -          &  55653.45418(1)       &  0.194958991(3)       &   3  &  18 \\
35 & DE CVn                    &  -                 &  M3         &  52784.05429(6)       &  0.364139237(9)       &  14  &  28 \\
36 & SDSS J132925.21+123025.4  &  -                 &  M6         &  55271.054831(4)      &  0.0809662425(5)      &  33  &  18 \\
37 & WD 1333+005               &  -                 &  M4.5       &  55611.476690(9)      &  0.121958759(1)       &  23  &  29,18,30 \\
38 & SDSS J134841.61+183410.5  &  CSS 21357         &  M3         &  56000.161920(8)      &  0.248431783(3)       &   9  &  18 \\
39 & QS Vir                    &  EC 13471-1258     &  M3         &  48689.1420(2)        &  0.150757475(4)       &  24  &  31 \\
40 & SDSS J141057.73-020236.7  &  CSS 07125         &  M3         &  53464.4888(1)        &  0.36349708(1)        &   7  &  18,5 \\
41 & SDSS J141126.20+200911.1  &  CSS 21055         &  T0         &  55991.388719(2)      &  0.0845327499(2)      &   9  &  32,33 \\
42 & SDSS J141134.70+102839.7  &  -                 &  M3         &  56031.1727(1)        &  0.16750971(4)        &   2  &  17,5 \\
43 & SDSS J141150.74+211750.0  &  -                 &  M3         &  55659.2477(1)        &  0.32163660(4)        &   4  &  1 \\
44 & GK Vir                    &  -                 &  M4.5       &  42543.33771(4)       &  0.344330839(1)       &  12  &  34,26 \\
45 & SDSS J142355.06+240924.3  &  CSS 080408        &  M5         &  55648.206115(5)      &  0.382004296(2)       &   7  &  18,5 \\
46 & SDSS J142427.69+112457.9  &  -                 &  -          &  54264.28247(2)       &  0.239293557(2)       &   5  &  1 \\
47 & SDSS J143547.87+373338.5  &  -                 &  M5         &  54148.2054(2)        &  0.125630956(9)       &  22  &  35,4,5 \\
48 & SDSS J145634.29+161137.7  &  CSS 09797         &  M6         &  51665.7893(30)  &  0.2291202(2)  &   2  &  18,5 \\
49 & SDSS J154057.27+370543.4  &  -                 &  M4         &  54913.4139(3)        &  0.26143556(5)        &   3  &  1 \\
50 & SDSS J154846.00+405728.7  &  -                 &  M6         &  54592.07303(2)       &  0.185515282(2)       &   4  &  4 \\
51 & NN Ser                    &  -                 &  M4         &  47344.02510(6)       &  0.130080129(1)       &  10  &  36,37 \\
52 & SDSS J164235.97-063439.7  &  -                 &  -          &  56770.19243(3)  &  0.28688831(49)  &   1  &  1 \\
53 & GALEX J171708.5+675712    &  -                 &  WD         &  55641.43159(7)       &  0.24613544(3)        &   1  &  38,39 \\
54 & RX J2130.6+4710           &  -                 &  M3.5       &  52785.1810(5)        &  0.52103658(6)        &  15  &  40 \\
55 & SDSS J220504.50-062248.6  &  -                 &  M2         &  54453.07812(7)       &  0.132386908(5)       &   9  &  1 \\
56 & SDSS J220823.66-011534.2  &  CSS 09704         &  M4         &  56175.879533(3)      &  0.156505699(2)       &   9  &  18 \\
57 & SDSS J223530.61+142855.0  &  -                 &  M4         &  55469.06504(9)       &  0.144456859(9)       &  10  &  17 \\
58 & SDSS J230627.54-055533.2  &  -                 &  -          &  55509.1090(7)  &  0.20008319(6)  &   1  &  1 \\
\hline
\end{tabular}
\end{table*}
\makeatother }

{ \footnotesize
\makeatletter
\begin{table*}
\contcaption{References for Table~\ref{tab:detachedtargets}.}
\begin{tabular}{llllllll}
\hline 
\multicolumn{8}{p{15cm}}{\scriptsize References: (1) \citet{Parsons15} - (2) Parsons et al. (in prep) - (3) \citet{Kleinman04} - (4) \citet{Pyrzas09} - (5) \citet{Rebassa-Mansergas12} - (6) \citet{Parsons12c} - (7) \citet{Law12} - (8) \citet{Bhatti10} - (9) \citet{Becker11} - (10) \citet{Parsons13b} - (11) \citet{Hoard07} - (12) \citet{Drake14b} - (13) \citet{Steinfadt10} - (14) \citet{Kaplan14} - (15) \citet{OBrien01} - (16) \citet{Maxted07} - (17) \citet{Parsons13} - (18) \citet{Drake10} - (19) \citet{Parsons12a} - (20) \citet{Drake14c} - (21) \citet{Parsons11} - (22) \citet{Bours15a} - (23) \citet{Irawati16} - (24) \citet{Pyrzas12} - (25) \citet{Nebot09} - (26) \citet{Parsons12b} - (27) \citet{Raymond03} - (28) \citet{vdBesselaar07} - (29) \citet{Farihi05} - (30) \citet{Farihi10} - (31) \citet{O'Donoghue03} - (32) \citet{Beuermann13} - (33) \citet{Littlefair14} - (34) \citet{Green78} - (35) \citet{Steinfadt08} - (36) \citet{Haefner89} - (37) \citet{Parsons10a} - (38) \citet{Vennes11} - (39) \citet{Hermes14c} - (40) \citet{Maxted04}}\\
\end{tabular}
\end{table*}
\makeatother }

{\footnotesize
\makeatletter
\begin{table*}
\caption{List of the 9 semi-detached eclipsing white dwarf binaries (cataclysmic variables) included in the monitoring programme described in this paper, sorted by RA. The SIMBAD identifier, frequently used alternative name and spectral type of the companion star to the white dwarf are given where available, with a star (*) indicating polars in which the white dwarf is strongly magnetic. The numbers in the parentheses following the zero-point and orbital period of the best linear ephemeris indicate the uncertainty in the last digits. The column marked with a \# shows the number of new eclipse times presented here and the last column lists references for the discovery, a detailed study of the binary and/or the companion's spectral type. }
\label{tab:cvtargets} 
\begin{tabular}{llllllll}
\hline 
ID & \multicolumn{2}{c}{white dwarf binary}  &  SpT$_2$       &  \multicolumn{2}{c}{best linear ephemeris}                 & \#  &  ref \\
   & SIMBAD identifier & alternative         &                &  T$_0$ (BMJD$_{\mathrm{TDB}}$) & P$_{\mathrm{orb}}$ (days) &     &      \\
\hline 
59 & HT Cas                    &  -                 &  M5.4       &  43727.4406(2)        &  0.073647180(1)       &  22  &  1,2,3 \\
60 & *FL Cet                   &  -                 &  M5.5       &  52968.82292(1)       &  0.0605163225(3)      &   9  &  4,5,6 \\
61 & SDSS J103533.02+055158.3  &  -                 &  BD         &  55353.952440(2)      &  0.05700667189(10)    &   8  &  7 \\
62 & NZ Boo                    &  -                 &  -          &  53799.14064(3)       &  0.0589094793(6)      &   6  &  8,9 \\
63 & SDSS J170213.24+322954.1  &  -                 &  M0         &  53647.73721(9)       &  0.100082204(3)       &   9  &  10,9,11 \\
64 & *V2301 Oph                &  -                 &  M5.5       &  48070.5244(2)        &  0.078449990(2)       &  17  &  12,13,3 \\
65 & EP Dra                    &  -                 &  -          &  47681.2295(2)        &  0.072656295(2)       &  10  &  14,15 \\
66 & V713 Cep                  &  -                 &  -          &  54337.87667(2)       &  0.0854185085(8)      &  11  &  16 \\
67 & *HU Aqr                   &  -                 &  M4.3       &  49102.4217(2)        &  0.086820371(2)       &  15  &  17,18,19,3 \\
\hline
\multicolumn{8}{p{15cm}}{\scriptsize References: (1) \citet{Patterson81} - (2) \citet{Feline05} - (3) \citet{Knigge06} - (4) \citet{Szkody02} - (5) \citet{O'Donoghue06} - (6) \citet{Schmidt05} - (7) \citet{Littlefair06} - (8) \citet{Szkody06} - (9) \citet{Savoury11} - (10) \citet{Szkody04} - (11) \citet{Littlefair06a} - (12) \citet{Barwig94} - (13) \citet{Ramsay07} - (14) \citet{Remillard91} - (15) \citet{Bridge03} - (16) \citet{Boyd11} - (17) \citet{Schwope93} - (18) \citet{Hakala93} - (19) \citet{Schwope11}}\\
\end{tabular}
\end{table*}
\makeatother }

\begin{table*}
\begin{center}
\caption{Telescopes and instruments used for eclipse observations, listed in alphabetical order.}
\label{tab:telescopes}
\begin{tabular}{ll}
\hline
telescope or &  details and/or explanation of acronym  \\
instrument   &                                         \\
\hline
ACAM         &  Imager mounted on the WHT. \\
DFOSC        &  Danish Faint Object Spectrograph and Camera, mounted on the DT. \\
DT           &  1.5 m Danish Telescope situated at La Silla, Chile. \\
HAWK-I       &  High-Acuity Wide-field K-band Imager on the VLT. \\
INT          &  2.5 m Isaac Newton Telescope, situated on La Palma, Spain. \\
LT           &  2.0 m robotic Liverpool Telescope, situated on La Palma, Spain.  \\
NTT          &  3.6 m New Technology Telescope situated at La Silla, Chile. \\
pt5m         &  0.5 m Durham/Sheffield telescope, situated on La Palma, Spain.  \\
RISE         &  High-speed photometer on the LT.  \\
SAAO         &  South-African Astronomical Observatory - 1m telescope + STE3 CCD camera \\
SOAR         &  4.1 m Southern Astrophysical Research telescope, situated at Cerro Pach\'on, Chile. \\
SOI          &  SOAR Optical Imager. \\
SOFI         &  Son of Isaac, infrared spectrograph and imaging camera on the NTT. \\
TNT          &  2.4 m Thai National Telescope, situated on Doi Inthanon, Thailand. \\
TRAPPIST     &  0.6 m robotic telescope at La Silla, Chile, equiped with TRAPPISTCAM photometer. \\
ULTRACAM     &  Three-channel high-speed photometer, \textbf{mounted on the WHT, NTT and VLT}.  \\
ULTRASPEC    &  High-speed photometer, \textbf{mounted on the TNT}.  \\
VLT          &  8.0 m Unit Telescope of the Very Large Telescope, situated on Paranal, Chile. \\
W1m          &  Warwick 1.0m telescope, situated on La Palma, Spain. \\
WFC          &  Wide-Field Camera on the INT. \\
WHT          &  4.2 m William Herschel Telescope, situated on La Palma, Spain. \\
\hline
\end{tabular}
\end{center}
\end{table*}


\section{The O-C method}
The method of timing a specific feature in the light curve of a star or binary and comparing this observed time to a time calculated from an ephemeris is a validated approach that has been in use for decades. The residuals in the form of observed minus calculated (O-C) times can be used to investigate the evolution of the star or binary in which the feature that is being timed originates. This technique is a powerful tool for revealing behaviour that deviates from the assumed model, as such deviations will show up as non-zero residuals, although the accuracy depends on the accuracy with which the feature itself can be timed. This feature is generally a steady periodic phenomenon, such as the pulses emitted by a rapidly-rotating neutron star \citep{Wolszczan92, Wolszczan94}, eclipses in binary stars \citep{Wood63, Parsons10b, Hermes12b, Lohr14}, or stellar pulsations \citep{Silvotti07, Mullally08, Hermes13a2}. Modelling these residuals can reveal the underlying process that causes them. Here, one can think of the presence of circumstellar or circumbinary planets, a change in orbital period due to angular momentum loss or redistribution through magnetic braking, gravitational wave emission or Applegate's mechanism. It might also be possible to detect long-term evolutionary processes such as white dwarf cooling which affects pulsation periods and amplitudes.

\begin{figure*}
\includegraphics{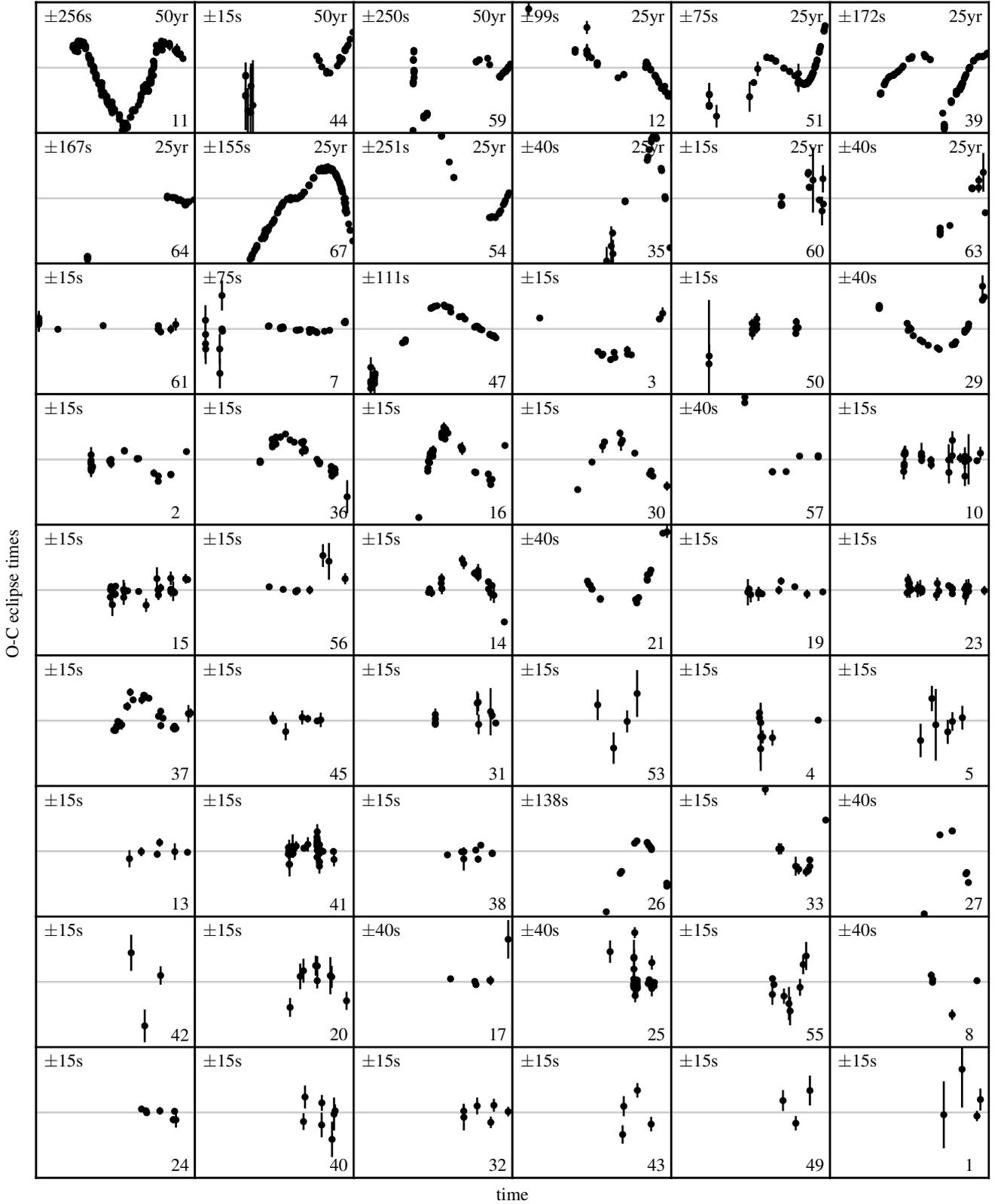}
\caption{Grid of O-C diagrams of 54 binaries in the eclipse timing programme, ordered by the
baseline of observations with the longest at the top left and the shortest at the bottom right. The horizontal axis of each thumbnail finishes at 1 July 2016 and spans 10~years, unless a longer baseline is indicated in the top right corner of the thumbnail. The number in the top left corner indicates the extent of the vertical axis, in seconds, for each O-C diagram. The number in the bottom right corner is the binary's ID number, corresponding to the ID numbers in Tables~\ref{tab:detachedtargets} and \ref{tab:cvtargets}. }
\label{fig:ocgrid}
\end{figure*}

In the remainder of this paper we will use this O-C method, applied to white dwarf eclipse times, to search for deviations from a constant orbital period. For this we assume a linear ephemeris that takes the form of 
\begin{equation}
T = T_0 + P_{\mathrm{orb}} \, E \,,
\end{equation}
for each binary. Here $P_{\mathrm{orb}}$ is the orbital period of the white dwarf binary, $T_0$ is the time at which the cycle number $E$ = 0, and $T$ is the time of a given orbital cycle $E$. The best linear ephemerides for the targets in the eclipse timing programme are listed in Tables~\ref{tab:detachedtargets} and~\ref{tab:cvtargets}, and were calculated using a linear least-squares approach to minimise the residuals. An overview of O-C variations of a \textbf{representative} sample of eclipsing white dwarf binaries is given in Fig.~\ref{fig:ocgrid}.

\section{Measuring eclipse times}
For each eclipse time we list the cycle number based on the best linear ephemeris, and the telescope+instrument combination used for obtaining the data, or the paper reference if the time has previously been published. As a representative example of the layout of the eclipse times tables, the eclipse times for SDSS\,J030308.35+005444.1 are shown in Table~\ref{tab:times:sdssj0303+0054}. Note that for data taken with ULTRACAM, which observes in three wavelength bands simultaneously, we present the weighted average of the three eclipse times. Also note that the times from \citet{Backhaus12} were published in the time standard BJD(TT). However, the difference between BJD(TT) and BJD(TDB) is at most 3.4\,ms, which is well within the uncertainties on the mid-eclipse times.

\begin{table*}
\caption{23 published and 11 unpublished mid-eclipse times for SDSS J0303+0054, a detached white dwarf + M-dwarf of spectral type M4.5 \citep{Rebassa-Mansergas12}. Numbers in parenthesis indicate the uncertainty in the last digit(s). Facilities used to obtain the new data include ULTRACAM on the \textbf{4.2m} William Herschel Telescope (WHT), RISE on the \textbf{2.0m} Liverpool Telescope (LT) and ULTRASPEC on the \textbf{2.4m} Thai National Telescope (TNT). }
\label{tab:times:sdssj0303+0054} 
\begin{tabular}{rlll}
\hline 
cycle     &  MJD(UTC)         &  BMJD(TDB)         &  source    \\
\hline
0         &  -                &  53991.11741(20)   &  \citealt{Pyrzas09}    \\
14        &  -                &  53992.99923(20)   &  \citealt{Pyrzas09}    \\
23        &  -                &  53994.20929(20)   &  \citealt{Pyrzas09}    \\
44        &  -                &  53997.03229(20)   &  \citealt{Pyrzas09}    \\
2559      &  -                &  54335.14302(20)   &  \citealt{Pyrzas09}    \\
2589      &  -                &  54339.17583(20)   &  \citealt{Pyrzas09}    \\
2960      &  -                &  54389.05324(20)   &  \citealt{Pyrzas09}    \\
2968      &  54390.122320(2)  &  54390.128292(2)   &  \citealt{Parsons10b}  \\
2976      &  54391.197780(2)  &  54391.203787(2)   &  \citealt{Parsons10b}  \\
3058      &  54402.221411(18) &  54402.227653(18)  &  \citealt{Parsons10b}  \\
11300     &  -                &  55510.262977(2)   &  \citealt{Parsons13b}  \\
11307     &  -                &  55511.204040(2)   &  \citealt{Parsons13b}  \\
11411     &  -                &  55525.185563(4)   &  \citealt{Parsons13b}  \\
13443     &  -                &  55798.362876(13)  &  \citealt{Backhaus12}  \\
13510     &  -                &  55807.370189(14)  &  \citealt{Backhaus12}  \\
13533     &  -                &  55810.462273(12)  &  \citealt{Backhaus12}  \\
13874     &  -                &  55856.305526(11)  &  \citealt{Backhaus12}  \\
13897     &  -                &  55859.397585(11)  &  \citealt{Backhaus12}  \\
13926     &  -                &  55863.296278(10)  &  \citealt{Backhaus12}  \\
13948     &  -                &  55866.253894(13)  &  \citealt{Backhaus12}  \\
16283     &  56180.161925(2)  &  56180.165824(2)   &  \citealt{Parsons13b}  \\
16505     &  56210.005261(2)  &  56210.010984(2)   &  \citealt{Parsons13b}  \\
16535     &  56214.038239(2)  &  56214.044118(2)   &  \citealt{Parsons13b}  \\
17011     &  56278.031604(17) &  56278.036452(17)  &  LT+RISE  \\
18671     &  56501.203048(30) &  56501.202982(30)  &  LT+RISE  \\
18701     &  56505.235763(5)  &  56505.236075(5)   &  WHT+ULTRACAM  \\
19191     &  56571.105019(13) &  56571.110566(13)  &  LT+RISE  \\
19800     &  56652.978910(13) &  56652.983082(13)  &  LT+RISE  \\
20035     &  56684.574551(5)  &  56684.575920(5)   &  TNT+ULTRASPEC  \\
21386     &  56866.201304(12) &  56866.201227(12)  &  LT+RISE  \\
22296     &  56988.533625(9)  &  56988.539519(9)   &  TNT+ULTRASPEC  \\
24941     &  57344.121062(10) &  57344.127255(10)  &  LT+RISE  \\
25108     &  57366.573012(5)  &  57366.578342(5)   &  TNT+ULTRASPEC  \\
25109     &  57366.707477(6)  &  57366.712799(6)   &  TNT+ULTRASPEC  \\
\hline 
\end{tabular}
\end{table*}

\subsection{Detached binaries}
\begin{figure}
\begin{center}
\includegraphics[width=0.5\textwidth]{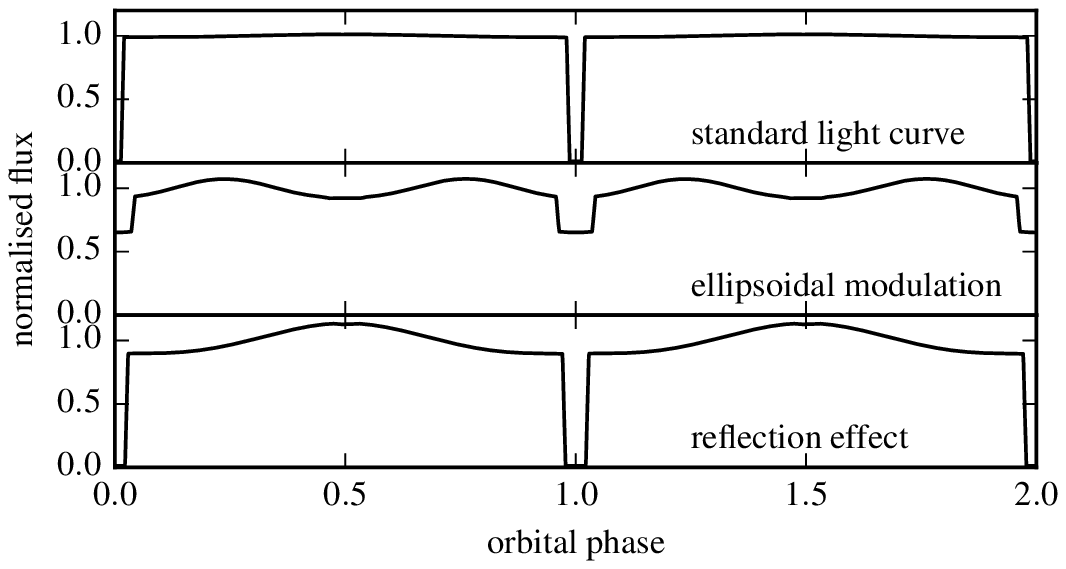}
\caption{Model light curves from \texttt{lcurve} of detached eclipsing WD+M-dwarf binaries with an inclination $i = 90^{\circ}$. \emph{Top:} simple detached binary. \emph{Middle:} light curve showing ellipsoidal modulation, caused by Roche lobe distortion of the secondary star. \emph{Bottom:} light curve showing the reflection effect, present in close binaries with a hot white dwarf. Because the stars are locked in synchronous rotation, one hemisphere of the M-dwarf will be strongly irradiated, and will therefore emit additional reprocessed radiation. Note that in all three types secondary eclipses at phase 0.5 are often not visible because the white dwarf covers such a small area of the M-dwarf, which is also much cooler. }
\label{fig:toymodellcs}
\end{center}
\end{figure}

To measure mid-eclipse times for the detached binaries we use the program \texttt{lcurve}\footnote{\raggedright The \texttt{lcurve} package was written by T.R.~Marsh; see \emph{http://www.warwick.ac.uk/go/trmarsh/software} for more information.}. This program is designed to model short-period white dwarf + main-sequence star binaries, and  can account for eclipses, deformation of the secondary star because it is close to filling its Roche lobe (ellipsoidal modulation), and reprocessed light from the white dwarf by the M-dwarf (reflection effect), see Fig.~\ref{fig:toymodellcs}. Limb darkening can be specified for both stars, using either a polynomial of up to fourth order, or the four-parameter law from \citet{Claret00}. Coefficients for the white dwarfs are taken from \citet{Gianninas13}, and those for late main-sequence stars from \citet{Claret11}. In addition it is possible to include gravitational lensing \citep{Marsh01} and the effect of gravity darkening for each star \citep{Claret11}. The latter becomes important for significantly Roche-distorted stars and for those stars that are rapidly rotating. Doppler beaming \citep{Loeb03} and a R{\o}mer delay \citep{Kaplan10} can be included as well, but these effects are generally negligible for the binaries in the eclipse timing programme. Some more details about \texttt{lcurve} can be found in \citet{Pyrzas09} and \citet{Copperwheat10}. For an example of a detailed study of an eclipsing binary that includes a reflection effect, ellipsoidal modulation, gravitational lensing and Doppler beaming, see \citet{Bloemen11}.

Once a good model is found for a binary, subsequent fits of new eclipse data only require as free parameters the mid-eclipse time $t_{\mathrm{mid}}$ and the secondary star's temperature $T_2$ to obtain a good fit, plus an overal linear or quadratic trend if significant changes \textbf{of the relative colours of the star} with airmass need to be modelled as well. By not optimising every parameter for every individual data set one avoids ending up with highly degenerate models and overly large uncertainty estimates.

\subsection{Semi-detached binaries} \label{sect:measuretimescvs}
Measuring eclipse times for semi-detached binaries is complicated by the inherent flickering in the light curves of these systems, which is caused by the varying accretion rate. On top of this, there are additional features near the white dwarf eclipse caused by the eclipse of the accretion disc and bright spot in non-magnetic systems and intermediate polars, and of the hot spot on the white dwarf's surface in polars. Therefore we chose to approach the fitting as done in \citet{Bours14b}, where the ingress and egress of the white dwarf eclipse are fit by a least-squares approach using a function that is composed of a sigmoid and a straight line,
\begin{equation}
y = \frac{k_1}{1 + e^{-k_2(x-k_3)}} + k_4 + k_5(x-k_3) \,.
\end{equation}
Here $x$ and $y$ are the time and flux measurements of the light curve, and $k_1$ to $k_5$ are coefficients of the fit. 

The straight line part allows fitting the overall trend outside and during ingress and egress. For polars, this includes the ingress and egress of the white dwarf itself, which can have a significant contribution, especially when the entire system is in a low state. The sigmoid part of the function fits the ingress or egress of the white dwarf, or the hot spot on the white dwarf for polars. To determine uncertainties, these fits are performed in a Monte Carlo manner in which the values of the data points are perturbed based on their uncertainties and the number of included data points are varied by a few at each edge, thereby reducing any strong effects in the results caused by single data points. 

Note that because of the presence of flickering and the varying mass-transfer rates the exact shape of the light curves of semi-detached binaries can vary significantly over a time scale of months - years. Measuring exactly the same feature in the white dwarf eclipse is therefore less straightforward than it is with detached binaries.


\section{Trends in eclipse time variations}
For some binaries it is possible to fit the eclipse times with models based on the assumption that circumbinary planets are present, but for most binaries such models have been refuted by additional data \citep{Parsons10b, Bours14b} or by detailed dynamical stability analyses \citep{Hinse12, Horner12b, Wittenmyer13}. It is somewhat more difficult to rule out the presence of Applegate's mechanism, although arguments on energetic grounds can be considered \citep{Brinkworth06,volschow16}. 

Because it can be difficult to confidently determine the cause of eclipse time variations in a given binary, we may be able to say something about the principal mechanism at work in these binaries by searching for trends in a large set of such systems. First of all, in Section~\ref{sect:baseline}, we explore whether a correlation exists between the amount of observed O-C variations and the baseline of the eclipse observations. Such a correlation would indicate that the present data set is still too limited to draw robust conclusions. In Section~\ref{sect:applegate}, we investigate the possibility that the O-C variations are the result of an Applegate-like mechanism. In particular, with increasing spectral type, from M0 to M8 and into the brown dwarf regime, such a mechanism is expected to become much less effective as the luminosity of the white dwarf's companion decreases. Therefore the O-C variations should become less pronounced, if not completely disappear. In addition, we expect to see no orbital period variations at all for double white dwarf binaries. This is because white dwarfs are not thought to experience magnetic cyclic behaviour, and because they have extremely small $R_2/a$ values. Therefore they cannot drive orbital period variations of the kind predicted by the Applegate and Lanza mechanisms \citep{Applegate92, Lanza98, Lanza06}.

In the event that the observed behaviour is caused by the presence of circumbinary planets, it is likely that there is no particular correlation present. This is because circumbinary planets can, in principle, form around a wide variety of binaries, and so there is no reason for them to be present preferentially around certain types of binaries. However, there may be a fundamental difference in the number and/or nature of the planets, depending on how planets form and/or survive around binary proto-stars and evolving close binaries \citep[see for example][]{Zorotovic13}. Either way, if circumbinary planets are present around the close white dwarf binaries such as those presented here, one has to be able to fit the O-C residuals with models of circumbinary planetary systems, which, in addition, have to be dynamically stable \citep{Marsh14}.

\subsection{Baseline of observations} \label{sect:baseline}
\begin{figure}
\begin{center}
\includegraphics[width=0.49\textwidth]{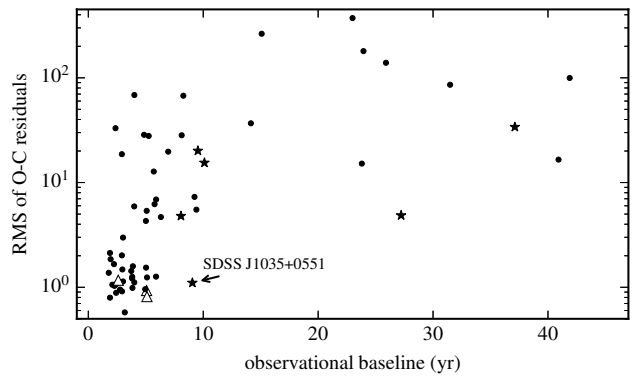}
\caption{Measure of the extent of eclipse timing variations with respect to a linear ephemeris (RMS) as a function of the baseline of the observations. Binaries with at least three eclipse time measurements have been included. The triangle, circle and star symbols represent detached double white dwarf binaries, other detached binaries and cataclysmic variables respectively. }
\label{fig:diffoc_end_logrms}
\end{center}
\end{figure}

An attempt to quantify the amount of eclipse timing variations as a function of the baseline of the observations for the various binaries is shown in Fig.~\ref{fig:diffoc_end_logrms}. Here, the root mean square (RMS) of the residuals is calculated using the standard formula,
\begin{equation}
\mathrm{RMS} = \sqrt{\frac{1}{N} \sum_{i} \Big( \frac{y_i - y(x_i)}{\sigma_i} \Big)^2 }\,,
\end{equation}
in which $x_i$, $y_i$ and $\sigma_i$ are the cycle number, eclipse time and uncertainty in the eclipse time and $y$ is the best linear ephemeris of these $N$ data points. Fig.~\ref{fig:diffoc_end_logrms} shows the RMS values for each binary as a function of the total baseline spanned by the eclipse observations for the given binary. 

It appears that the RMS saturates at a value near 100, although the log-scale of the plot enhances this feature. Nevertheless, the figure indicates that any white dwarf + low-mass main sequence star binary with eclipse observations spanning at least 10~years is extremely likely to show significant residuals in the O-C eclipse times. The only real exception in our sample so far is SDSS\,J1035+0551, which is a cataclysmic variable with a brown dwarf donor \citep{Littlefair06}. The observational baseline is close to 10~yrs for this binary, and the O-C times are perfectly flat (see the thumbnail with ID number 61 in Fig.~\ref{fig:ocgrid}).

\begin{figure}
\begin{center}
\includegraphics[width=0.49\textwidth]{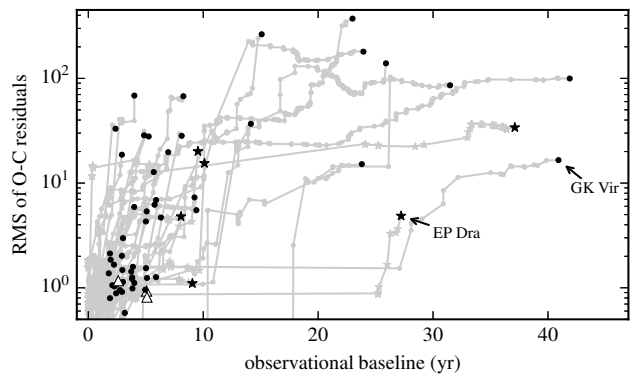}
\caption{Measure of the extent of eclipse timing variations with respect to a linear ephemeris (RMS) as a function of the baseline of the observations, starting with the first three eclipse times and subsequently increasing the number of data points included by one for the next RMS calculation. See the text for more details. The triangle, circle and star symbols represent detached double white dwarf binaries, other detached binaries and cataclysmic variables respectively. }
\label{fig:diffoc_end_logrms_grey}
\end{center}
\end{figure}

This idea is reinforced by Fig.~\ref{fig:diffoc_end_logrms_grey}, which, in addition to the points in Fig.~\ref{fig:diffoc_end_logrms}, also shows the RMS values of intermediate sets of eclipse times in grey. These intermediate RMS values are calculated using an integer number of eclipse times, starting with the first three, increasing by one with each step, and calculating the best linear ephemeris and corresponding RMS for these sets of eclipse times. One of two clear exceptions to the general trend is shown by GK\,Vir, which continues at an RMS close to 1, until the baseline of the observations reaches $\sim$~27~years. However, this behaviour is caused by an extremely large gap of nearly the same duration in the eclipse observations, rather than by an actual feature of the data. The second exception, EP\,Dra, has a similar gap in the observations, in this case of roughly 22~years.

It appears that the baseline of the eclipse observations is indeed quite an important factor in determining whether or not O-C variations are present. Although not unexpected, the long minimum baseline of $\sim$~10~years required means that a lot of data needs to be acquired before robust general conclusions can be drawn. This is particularly important because mixing short and long baselines may work to obscure trends in the data.

\subsection{Are the O-C variations caused by magnetic cycles in the secondary stars?} \label{sect:applegate}
It is possible that the observed O-C variations are caused by the presence of magnetic cycles in the secondary stars, and therefore represent true variations in the orbital periods of these binaries. Because Applegate's mechanism is driven by magnetic cycles in the companion, and because magnetic activity decreases towards later spectral types, a correlation between the spectral type and the amount of O-C variations is to be expected. In addition, the maximum energy available for driving the magnetic cycles is given by the star's total luminosity, which correlates steeply with the star's mass and spectral type. For binaries with later spectral types, and masses close to the stellar -- brown dwarf limit, we therefore expect much smaller O-C variations, if any are present at all.

\begin{figure}
\begin{center}
\includegraphics[width=0.49\textwidth]{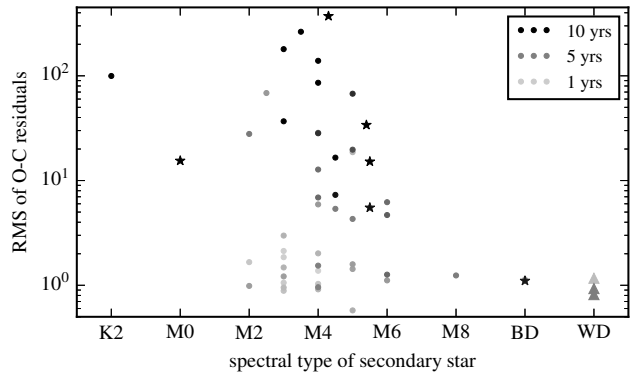}
\caption[RMS of O-C variations as a function of secondary's spectral type.]{Measure of the extent of eclipse timing variations with respect to the best linear ephemeris (RMS) as a function of the secondary star's spectral type. The grey scale corresponds to the baseline of the eclipse observations, with white for 0~yrs and black for 10~yrs as indicated in the legend. All binaries with a baseline exceeding 10~yrs are plotted in black to avoid excessive stretching of the grey scale. The triangle, circle and star symbols represent detached double white dwarf binaries, other detached binaries and cataclysmic variables respectively. \textbf{Note that the RMS values for the two binaries in our sample with brown dwarf companions are very similar, so that their symbols overlap.} }
\label{fig:diffoc_end_logrms_spt}
\end{center}
\end{figure}

To search for such a correlation, Fig.~\ref{fig:diffoc_end_logrms_spt} shows the RMS of the eclipse time residuals with respect to the best linear ephemeris as a function of the secondary star's spectral type. In addition, the grey scale indicates the length of the baseline of the eclipse time observations, which continues to be an important factor. 

From Fig.~\ref{fig:diffoc_end_logrms_spt} it appears that there could indeed be a correlation between the amount of O-C variations and the spectral type of the secondary star. However, note that we do not yet have binaries with baselines exceeding 10~yrs at all secondary star spectral types. Nonetheless, binaries with secondaries of spectral type M6 -- M8, brown dwarfs or white dwarfs evidently have smaller RMS values. Our sample of binaries with such late-type secondaries is limited to 11. Of these, SDSS\,J103533.02+055158.3, a cataclysmic variable with a brown dwarf donor \citep{Littlefair06}, has the longest observational baseline of 9~yrs. Note that when we only consider binaries with baselines smaller than 10~yrs, the RMS values of binaries with secondary star spectral types earlier than M6 are on average still larger than those of binaries with later spectral types. It therefore appears that there are indeed two different populations, with the separating line close to spectral type M6. This behaviour may result from the transition of low-mass main sequence stars having a radiative envelope and convective core at earlier spectral types, to being fully convective at late spectral types \citep{Chabrier97}. Due to this absence of the tachocline in late-type stars, the magnetic fields may be generated through a different mechanism than in early-type stars \citep{Morin08,Morin10}. Studying which secondary stars drive orbital period variations in white dwarf binaries can therefore shed light on the presence and generation of magnetic dynamos in low-mass main sequence stars. In addition, long term monitoring can reveal the temporal variability of such magnetic fields through the observed orbital period variations.

Ideally, we would like to investigate whether the RMS of the O-C variations correlate with a parameter that represents the strength of an Applegate-like mechanism in a given secondary star. This parameter would primarily depend on the secondary star's luminosity, or equivalently, its mass \citep{Applegate92,volschow16}. The reason is that this parameter is representative of the maximum energy available in the star for driving the mechanism. A second important parameter would be the binary's orbital separation, representing the ease with which the mechanism can couple to the binary orbit. For larger separations, one needs a larger variation in the star's gravitational quadrupole moment in order to obtain orbital period variations of the same magnitude. Assuming an essentially constant period for the magnetic cyles as well as a constant period change relative to the binary's orbital period, the energy required to drive the O-C variations, $\Delta E$, is given by
\begin{equation}
\frac{\Delta E}{E_{\mathrm{sec}}} \propto a_{\mathrm{bin}}^2 M_{\mathrm{sec}}^2 R_{\mathrm{sec}}^{-3} L_{\mathrm{sec}}^{-1} \propto a_{\mathrm{bin}}^2 M_{\mathrm{sec}}^{-3.45}
\end{equation}
with $E_{\mathrm{sec}}$, $M_{\mathrm{sec}}$, $R_{\mathrm{sec}}$ and $L_{\mathrm{sec}}$ the energy, mass, radius and luminosity of the secondary star, and $a_{\mathrm{bin}}$ the binary's orbital separation \citep{volschow16}.

However, for most binaries in our eclipse timing programme we do not know the masses of the secondary stars or the orbital separations of the binaries, and we only have the spectral types of the secondary stars. Estimates of masses from spectral types are typically quite inaccurate given that we do not know the age of the binaries, especially for late-type M-dwarfs. Therefore, because the star's luminosity and spectral type are closely related, until we know more about the individual binaries and secondary stars, we use the spectral types as an indicator for magnetic activity (see Tables~\ref{tab:detachedtargets} and \ref{tab:cvtargets}).

\section{A look at a few selected binaries}
This section includes a more detailed look at a few selected binaries, which are either interesting in their own right, or show behaviour representative of a group of targets in our timing programme.

\subsection{Binaries with long observational baselines}
First of all we have a number of binaries for which the observational baseline is long, and which all show substantial O-C variations. As a member of this category, RX\,J2130.6+4710 is particularly interesting because this is the first time that its O-C variations have been detected, even though they have a large amplitude. Subsequently, we briefly discuss the observations of NN\,Ser and QS\,Vir, both of which have long histories concerning their O-C variations and proposed models to explain them.

\subsubsection{RX\,J2130.6+4710}
RX\,J2130.6+4710 is a detached white dwarf + M-dwarf binary, with an orbital period of $\sim$~12.5 hours. A study of the system parameters based on phase-resolved spectroscopy and ULTRACAM photometry of this binary was published by \citet{Maxted04}, who also published the first mid-eclipse times. Note that the eclipse times they measured from data using the \textbf{1.0m} Jakobus Kapteyn and \textbf{2.5m} Isaac Newton Telescopes on La Palma cannot be trusted to better than several seconds due to inaccurate timestamping of the data. We have therefore set the uncertainties of these times to 10~s. Unfortunately, RX\,J2130.6+4710 lies only 12\arcsec~away from a bright G0 star (HD\,204906), which complicates the extraction of the light curve, especially with data taken during variable atmospheric conditions. In addition, the M3.5 -- M4 main-sequence star \citep{Maxted04} frequently experiences flares. 

\begin{figure}
\begin{center}
\includegraphics[width=0.5\textwidth]{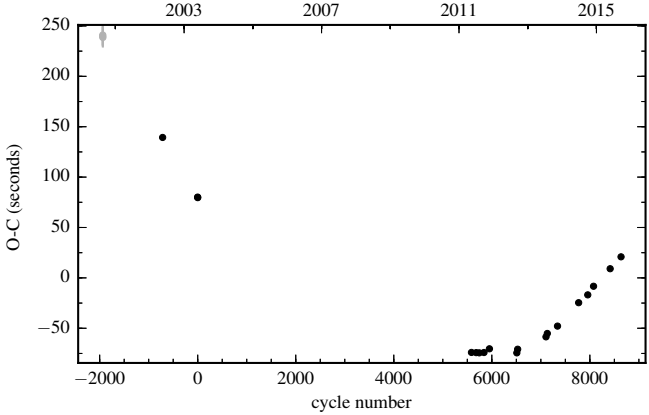}
\caption{O-C diagram of RX\,J2130.6+4710, with respect to the best linear ephemeris in Table~\ref{tab:detachedtargets}. Note that the three eclipse times near cycle number -2000 are not reliable to within several seconds due to inaccurate timestamping of the data \citep{Maxted04}. Times with uncertainties larger than 3~seconds are shown in grey. }
\label{fig:oc_rxj2130.6+4710}
\end{center}
\end{figure}

We obtained 15 new eclipse times with ULTRACAM, ULTRASPEC and RISE. These new times reveal an extreme deviation in the mid-eclipse times with respect to a linear ephemeris, see Fig.~\ref{fig:oc_rxj2130.6+4710}, to the extent that the eclipse in September 2015 was observed almost 11~minutes later than expected from the original ephemeris published by \citet{Maxted04}. The overall shape of the O-C times appears to be parabolic-like, and could correspond to part of a sinusoidal variation. Such a sinusoidal variability could result from an Applegate-like mechanism, or be the result of a reflex motion of the binary caused by a third companion. Given our observational baseline of 15~yrs, the mechanism at work operates with a period exceeding at least 30~years. Explaining the large amplitude of the O-C measurements with the presence of a third object would require a brown dwarf companion. Currently, the data covers too small a section of such a sinusoid to constrain the parameters of any sinusoidal fit. Regular eclipse observations in the coming years are necessary to shed further light on the O-C variations in this binary.

\subsubsection{NN\,Ser}
NN\,Ser is a white dwarf + M-dwarf binary with an orbital period of 3.1~h, showing not only eclipses but also a large reflection effect. This system has been studied extensively, both photometrically and spectroscopically \citep[see][]{Haefner89, WoodJ91, Catalan94, Haefner04, Parsons10a}. About 10 years ago, \citet{Brinkworth06} noticed that NN\,Ser appears to show a decrease in its orbital period, and that this decrease is much larger than predicted by models of magnetic braking in close binaries or by the mechanism proposed by \citet{Applegate92}. Subsequent changes in the sign of the O-C variations definitively ruled out magnetic braking as the cause. Since then, the changes in the mid-eclipse times have been \textbf{attributed to the presence of circumbinary substellar companions \citep{Qian09}. A more recent model includes} two circumbinary Jovian planets, \textbf{and} has been refined with each release of new eclipse times \citep{Beuermann10,Beuermann13,Marsh14}. Out of all the white dwarf binaries showing eclipse timing variations, NN\,Ser is the only one in which the proposed circumbinary planetary models have survived the addition of new eclipse times, as well as rigorous dynamical stability analysis. Even the O-C variations in the secondary eclipse times follow the proposed model, thereby ruling out apsidal precession as the cause of the observed variations \citep{Parsons14}. \textbf{A detection of a dust disc around NN~Ser further supports the idea that circumbinary planets can form and exist around evolved binaries \citep{Hardy16}.}

\begin{figure}
\begin{center}
\includegraphics[width=0.45\textwidth]{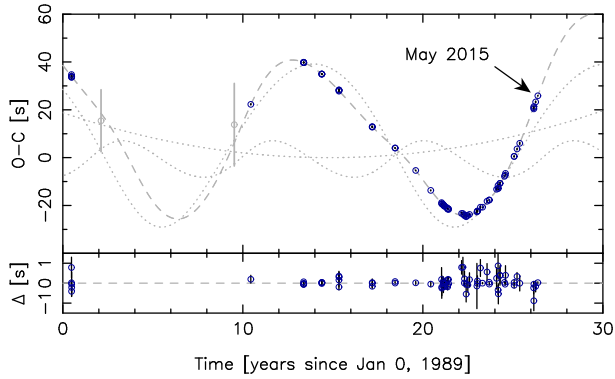}
\caption{O-C diagram of NN\,Ser. The dashed line indicates the best fit model, which includes two circumbinary Jovian planets as well as a quadratic trend with a much longer period (dotted lines). The lower panel shows the residuals with respect to the best fit. }
\label{fig:oc_nnser}
\end{center}
\end{figure}

\begin{figure}
\begin{center}
\includegraphics[width=0.45\textwidth]{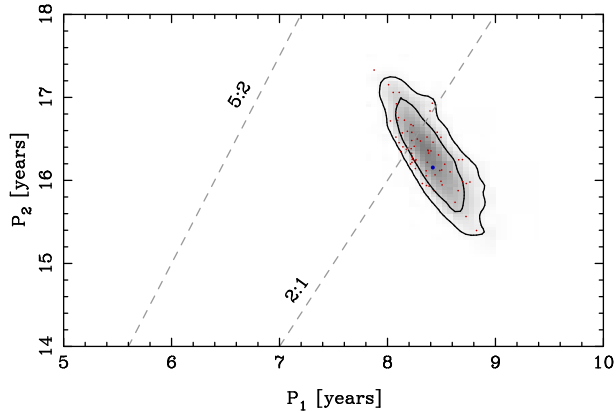}
\caption{Parameter space showing the periods of the two circumbinary Jovian planets proposed in the model for NN\,Ser. The red dots indicate a dynamical stability lasting 0.1 -- 1~Myr, while the single, larger, blue dot indicates a model that is dynamically stable for a time exceeding 1~Myr. Compare with the right-most panel of Fig.~5 in \citet{Marsh14}. }
\label{fig:nnser_params}
\end{center}
\end{figure}

\begin{figure}
\begin{center}
\includegraphics[width=0.45\textwidth]{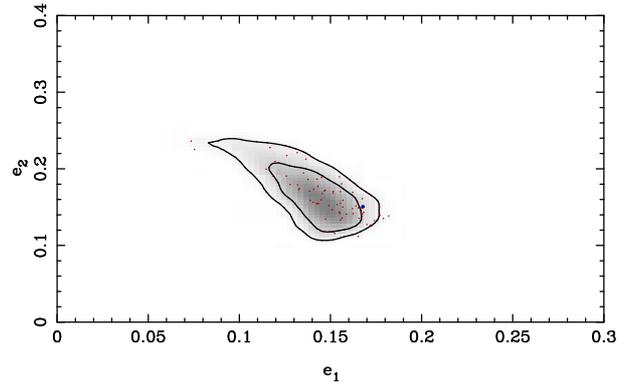}
\caption{Parameter space showing the eccentricities of the two circumbinary Jovian planets proposed in the model for NN\,Ser. The red dots indicate a dynamical stability lasting 0.1 -- 1~Myr, while the single, larger, blue dot indicates a model that is dynamically stable for a time exceeding 1~Myr. Compare with Fig.~6 in \citet{Marsh14}. }
\label{fig:nnser_e1e2}
\end{center}
\end{figure}

As part of the eclipse timing programme, we present 10 new eclipse times of NN\,Ser. With these new times, we have recalculated the planetary model fits and dynamical stability analyses presented in \citet{Marsh14}. In order to obtain a good fit, the model now requires a non-zero quadratic term in the ephemeris, therefore corresponding to model `B + $e_2$ + $\beta$' from \citet{Marsh14}. The best model, together with the O-C eclipse times, is shown in Fig~\ref{fig:oc_nnser}. The number of dynamically stable models has decreased significantly, leaving only 1 model in 79700 that is stable for more than 1~Myr (the age of the close binary itself), see Figs.~\ref{fig:nnser_params} and \ref{fig:nnser_e1e2}. 

The periods of the two Jovian planets are further constrained by the additional data and now definitively favour a period ratio close to 2:1. The models with planetary periods close to the 5:2 ratio \citep{Beuermann10,Marsh14} are no longer viable. \textbf{The eccentricities of the planetary orbits are both non-zero, and the distribution of these parameters is shown in Fig.~\ref{fig:nnser_e1e2}.} The long-period quadratic term in the model is in the direction of a lengthening orbital period, and can therefore not be explained by natural processes that lead to angular momentum loss, such as magnetic braking or gravitational wave emission. Although it is far too early to say anything definitive about its origin, the quadratic term could be attributed to a third, more distant, circumbinary object. Note however that, given the spectral type M4 of the secondary star \citep{Parsons10a}, some form of Applegate's mechanism could well be at work in this binary. If both Applegate's mechanism and circumbinary planets are present, this severely complicates the process of modelling the eclipse times, since we cannot model the effect of the magnetic activity cycle. The best test of the planetary model with two Jovian planets will come in 2018-2019, when the model predicts a maximum and subsequent downturn in the O-C eclipse times.

\subsubsection{QS\,Vir}
QS\,Vir is a white dwarf + M-dwarf binary, also known as EC\,13471-1258, discovered in the Edinburgh-Cape blue object survey \citep{Stobie97}. The red dwarf has a spectral type of M3, and almost completely fills its Roche lobe \citep{O'Donoghue03}, and the binary has therefore been classified as a hibernating cataclysmic variable by those authors. However, analysis of the white dwarf rotation showed that the system could also be a pre-cataclysmic variable \citep{Parsons11a,Parsons16a}, although the hibernation theory is not fully excluded \citep{DrakeJ14}. Recently, it was also discovered that prominences from the M-dwarf appear to be locked in stable configurations within the binary system, and last there for more than a year \citep{Parsons16a}. Besides the white dwarf eclipse, the binary's light curve shows a small reflection effect at blue wavelengths, and ellipsoidal modulation at redder wavelengths \citep{Parsons10b}. 

\begin{figure}
\begin{center}
\includegraphics[width=0.5\textwidth]{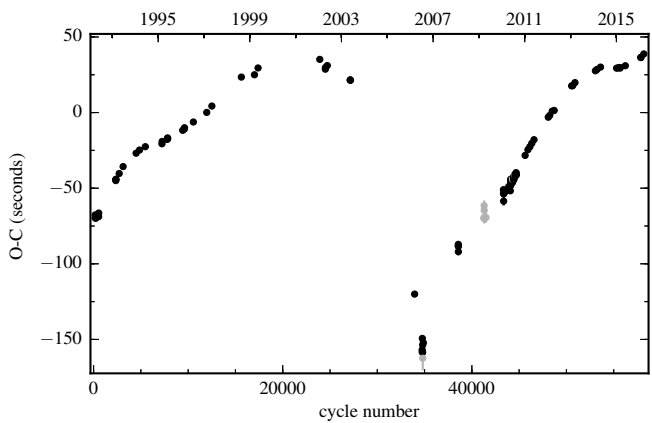}
\caption{O-C diagram of QS\,Vir, with respect to the linear ephemeris in Table~\ref{tab:detachedtargets}. Times with uncertainties larger than 3~seconds are shown in grey. }
\label{fig:oc_qsvir}
\end{center}
\end{figure}

There are 86 published mid-eclipse times as well as 24 unpublished as part of the eclipse timing programme presented here. As can be seen in Fig.~\ref{fig:oc_qsvir}, QS\,Vir shows eclipse time variations with large amplitudes and with occasional extreme changes. From the latest eclipse times it appears that another local or absolute maximum might have been reached in the O-C residuals, similar perhaps to the O-C variations close to cycle number 5000 or 20000. Observations in the coming years will show if another abrupt shift occurs, and therefore whether or not the O-C variations are cyclic in their behaviour.

There have been several attempts to explain the cause of these large and erratic O-C variations. \citet{Qian10}, following \citet{Brinkworth06}, calculated the energy available in the secondary star, and showed that this was insufficient to cause the observed large-amplitude O-C variations through Applegate's mechanism. Instead, they proposed a combination of a large continuous decrease in the binary's orbital period and the presence of a circumbinary planet of $\sim$~7~M$_{\mathrm{Jup}}$. New eclipse data quickly showed that this hypothesis was wrong \citep{Parsons10b}. \citet{Almeida11} then presented a new fit to the data, which included two circumbinary planets. However, the extreme shift near cycle number 30000 forces at least one planet into a highly eccentric orbit, causing the entire planetary system to be dynamically unstable \citep{Horner13}. 

Clearly, the eclipse time variations in this binary are complex, and what causes them remains to be discovered. There are a few other binaries in the timing programme that show similarly large O-C variations, such as V471\,Tau \citep{Hardy15} and HU\,Aqr \citep{Bours14b}.

\subsection{Binaries with little or no O-C variability}
The binaries mentioned above have eclipse observations spanning over at least a decade, but there are many binaries in the eclipse timing programme that have a shorter baseline. Some of these already show small O-C variations, while the data for other systems is still consistent with a constant orbital period.

\subsubsection{SDSS\,J090812.03+060421.2, aka CSS\,080502}
\begin{figure}
\begin{center}
\includegraphics[width=0.5\textwidth]{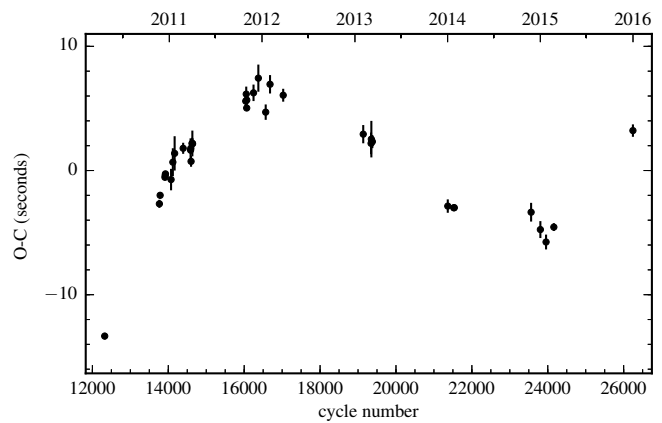}
\caption{O-C diagram of CSS\,080502, also known as SDSS\,J090812.03+060421.2, with respect to the linear ephemeris in Table~\ref{tab:detachedtargets}. }
\label{fig:css080502}
\end{center}
\end{figure}

CSS\,080502 is a detached white dwarf + M-dwarf binary with an observational baseline of about five years, and it has already started to show some O-C variations, see Fig~\ref{fig:css080502}. It was discovered as an eclipsing white dwarf binary in CSS data \citep{Drake09}, and has also been observed as part of SDSS as SDSS\,J090812.03+060421.2. \citet{Pyrzas09} determined approximate parameters for the white dwarf and M-dwarf through decomposition and fitting of the available SDSS spectra. The M-dwarf has a spectral type of M4, as determined by \citet{Rebassa-Mansergas12}, which is in good agreement with the previous determinations of \citet{Drake10} and \citet{Silvestri06}.

CSS\,080502 is a representative example of a number of targets in the timing programme, all of which have eclipse observations covering a few years and have started to show small-scale O-C variations on the order of $\pm$ 5 -- 10~seconds. Among others, this includes SDSS\,J1210+3347 and SDSS\,J1212-0123.

\subsubsection{SDSS\,J093947.95+325807.3, aka CSS\,38094}
There are also a few binaries which have so far shown no variability in their eclipse arrival times. One example of this class is CSS\,38094, also known as SDSS\,J093947.95+325807.3. CSS\,38094 was discovered in CSS data as a white dwarf + red dwarf binary, with the latter having a spectral type of M5. The more recent determination of spectral type from SDSS data agrees with this \citep{Rebassa-Mansergas12}. The first white dwarf eclipse times were published by \citet{Backhaus12}, and we obtained 6 more as part of our timing programme. 

\begin{figure}
\begin{center}
\includegraphics[width=0.5\textwidth]{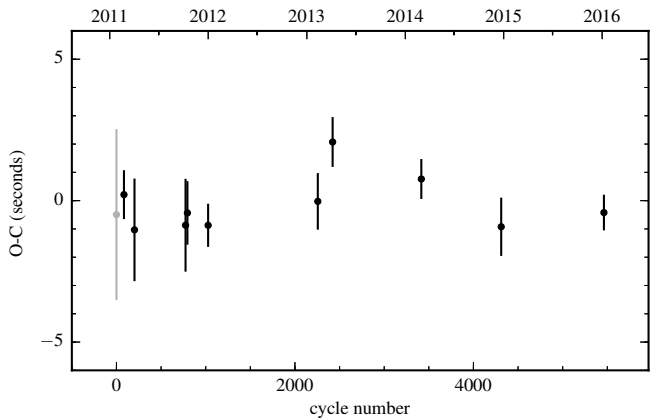}
\caption{O-C diagram of CSS\,38094, also known as SDSS\,J093947.95+325807.3, with respect to the linear ephemeris in Table~\ref{tab:detachedtargets}. Times with uncertainties larger than 3~seconds are shown in grey. }
\label{fig:css38094}
\end{center}
\end{figure}

The O-C diagram of CSS\,38094 is shown in Fig.~\ref{fig:css38094}, and is so far consistent with a linear ephemeris. Binaries with similarly flat O-C diagrams are SDSS\,J0314+0206, CSS\,080408 and CSS\,03170. However, for all of these binaries the eclipse observations span only a few years, and it is therefore too early to say whether these binaries indeed show no variations at all in the eclipse arrival times.

\subsection{Two long-period binaries}
We have two white dwarf + M-dwarf binaries with \textbf{relatively} long orbital periods in our eclipse timing programme. These are SDSS\,J092741.73+332959.1, with a period of 2.3~days and an observational baseline of $\sim$~3~yrs, and SDSS\,J095737.59+300136.5, with an orbital period of 1.9~days and so far only one observed eclipse. Because of the large separation between the two stars in these binaries, we expect any Applegate-type mechanism to couple extremely weakly to the binary orbit. In this case, the O-C residuals should be consistent with zero. So far, the baselines are too short to draw conclusions.

In the long term, these two targets are prime indicators of whether or not an Applegate- or Lanza-type mechanism operates in white dwarf + M-dwarf binaries. Note however, that the long orbital periods also complicate the observations, primarily because the periods are close to a multiple of 24~h.

\subsection{Binaries with brown dwarf or white dwarf secondaries}
There are three double white dwarf binaries in our programme, CSS\,41177 \citep{Parsons11,Bours14a,Bours15a}, GALEX\,J1717+6757 \citep{Vennes11,Hermes14c} and NLTT\,11748 \citep{Steinfadt10,Kawka10,Kilic10b,Kaplan14}. There are also two white dwarf + brown dwarf binaries in the eclipse timing programme, CSS\,21055 \citep{Beuermann13b,Littlefair14} and SDSS\,J103533.02+055158.3 \citep{Littlefair06,Savoury11}. 

Note that for NLTT\,11748, the eclipse times published by \citet{Kaplan14} were said to be corrected to the solar system barycenter. However, together with those authors, we discovered that they were in fact converted to Barycentric Coordinate Time (TCB) and not Barycentric Dynamical Time (TDB). These two are linearly related\footnote{\raggedright See the IAU resolution at www.iau.org/static/resolutions/IAU2006\_Resol3.pdf.}, and we have used this to convert the times from \citet{Kaplan14} to BMJD(TDB) in order to be consistent with the new eclipse times measured as part of our eclipse timing programme. 

As we do not expect any strong magnetic cyclic activity in white dwarfs or brown dwarfs, we expect the O-C diagrams of these binaries to be flat, consistent with a constant orbital period. The observational baselines for these five binaries range from 3~yrs to 9~yrs and are indeed all consistent with constant orbital periods.


\section{Conclusions}
The large eclipse timing programme presented here is the first step towards revealing the extent and amplitude of eclipse timing variations throughout the class of white dwarf binaries. In addition, it will enable the systematic search for correlations between the amount of eclipse timing variability and characteristics of the systems, such as the secondary star's spectral type. 

Currently, we are mostly limited by the relatively short observational baselines for the majority of our targets. However, there are some preliminary conclusions that we can draw at this point.

\begin{itemize}
	\item All white dwarf + M-dwarf binaries in the eclipse timing programme with observational baselines exceeding 10~yrs show O-C residuals on the order of 100 seconds. 
	\item It appears that the presence of a circumbinary planetary system can be ruled out for almost all of these binaries with long baselines and large O-C variations. The one exception is the white dwarf + M-dwarf binary NN\,Ser, for which the data can still be fit with a model including two Jovian planets and a quadratic term in the ephemeris.
	\item Our programme contains 11 white dwarf binaries with secondary stars of spectral types M6 -- M8, brown dwarfs or white dwarfs. \textbf{Only} two of these have an RMS $>$~1.3 for the O-C residuals. The first is SDSS\,J0106-0014, which has an M-dwarf with spectral type M6 \citep{Rebassa-Mansergas12}, and shows O-C variations of $\pm$ 5~s over a baseline of $\sim$ 6~years. The other binary, SDSS\,J1329+1230, has a very similar baseline, and also shows variations on the $\pm$ 5~s scale. The spectral type of the M-dwarf in this binary is somewhat uncertain, because it is not visible spectroscopically at optical wavelengths. As a best guess, it has been classified as an M6 from the SDSS spectra \citep{Drake10}, but it could well be a subtype or two earlier or later.
 	\item There is currently only one close eclipsing binary that has a long observational baseline of eclipse observations which are consistent with a constant orbital period, namely AA\,Dor \citep{Kilkenny11, Kilkenny14}. This binary contains a hot subdwarf OB star and a low-mass companion at the substellar limit \citep{Vuckovic16}. Its observational baseline is now about 37~yrs \citep{Lohr14,Kilkenny14}. Given the low mass of the secondary star, this is in full agreement with the small RMS values that we found here for the white dwarf binaries with low mass secondaries.
\end{itemize}

Overall, we believe that it is most likely that the observed orbital period variations originate in the secondary stars in these binaries and work on the binary orbit through an Applegate- or Lanza-type mechanism. The fact that Applegate's mechanism can apparently be ruled out for some binaries with extreme O-C variations could be due to the fact that we do not completely understand either the processes acting in the low-mass main sequence stars or the way these couple to the binary orbits. Also, \citet{Lanza98} and \citet{Lanza06} have suggested that the coupling of magnetic cycles to the orbital period could occur using only a fraction of the energy of the mechanism suggested by \citet{Applegate92}. In the future, with an expanded version of the timing programme presented here, we may be able to understand and possibly calibrate the magnetic behaviour of low-mass main-sequence stars. This could be of crucial importance for understanding planetary systems around such single low-mass stars, as well as observed transit-timing variations of, for example, hot-Jupiters around such stars \citep[see for example][]{Watson10,Maciejewski16}. 

Finally, we expect that in practice the magnetic mechanism is entangled with other phenomena present in the white dwarf binaries studied here. This could include strong irradiation and subsequent inflation of the secondary star which could redistribute angular momentum, or the presence of a weak magnetic field in the white dwarf and the interaction between the stellar magnetic fields. Given the prevalence of planetary systems in all kinds of configurations as well as the fact that a significant fraction of white dwarfs continuously appear to be accreting heavy metals \citep{Gaensicke12,Koester14}, it seems reasonable to assume that some planetary systems -- or possibly their remnants -- do exist around evolved white dwarf binaries. This is exactly what might be causing the O-C variations in NN\,Ser. However, we reiterate that for the vast majority of systems the data are inconsistent with O-C variations caused by external, line-of-sight variations.


\section*{Acknowledgements}
MCPB acknowledges support from the Joint Committee ESO-Government of Chile (grant 2014). 
TRM, EB, VSD, SPL, TB and RWW are grateful to the Science and Technology Facilities Council for financial support in the form of grants \#ST/L000733, \#ST/L00075X/1 and \#ST/M001350/1.
SGP and CC acknowledge support from CONICYT FONDECYT grants 3140585 and 3140592 respectively. CC also acknowledges support from the Millennium Nucleus RC130007 (Chilean Ministry of Economy).
JJH acknowledges support from NASA through Hubble Fellowship grant \#HST-HF2-51357.001-A, awarded by the Space Telescope Science Institute, which is operated by the Association of Universities for Research in Astronomy, Incorporated, under NASA contract NAS5-26555.
DK is partially funded by the University of the Western Cape and the National Research Foundation (NRF) of South Africa.










\bsp	
\label{lastpage}
\end{document}